\documentstyle[aps,epsf,psfig,twocolumn]{revtex}
\begin{document}
\draft
\flushbottom
\twocolumn[
\hsize\textwidth\columnwidth\hsize\csname @twocolumnfalse\endcsname

\title{
Doping- and size-dependent suppression of tunneling in carbon nanotubes}
\author{S. Bellucci $^1$, J. Gonz\'alez $^2$ and P. Onorato $^1$ $^3$\\}
\address{
        $^1$INFN, Laboratori Nazionali di Frascati,
        P.O. Box 13, 00044 Frascati, Italy. \\
        $^2$Instituto de Estructura de la Materia,
        Consejo Superior de Investigaciones Cient{\'\i}ficas,
        Serrano 123, 28006 Madrid, Spain.
        $^3$Dipartimento di Scienze Fisiche,
        Universit\`{a} degli Studi di Napoli ``Federico II'',
        Via Cintia, I-80126 Napoli, Italy.}
\date{\today}
\maketitle
\widetext
\begin{abstract}
We study the effect of doping in the suppression of tunneling observed
in multi-walled nanotubes, incorporating as well the influence of the
finite dimensions of the system. A scaling approach allows us to
encompass the different values of the critical exponent $\alpha $
measured for the tunneling density of states in carbon nanotubes.
We predict that further reduction of $\alpha $ should be observed in
multi-walled nanotubes with a sizeable amount of doping. In the case of
nanotubes with a very large radius, we find a pronounced crossover between
a high-energy regime with persistent quasiparticles and a low-energy
regime with the properties of a one-dimensional conductor.
\end{abstract}


]
\narrowtext
\tightenlines

During recent years there has been much interest in the
investigation of the electronic properties of carbon nanotubes
(CN)\cite{1}. The reduced dimensionality of these systems leads to
the appearance of unconventional effects, such as a suppression of
the tunneling conductance at low energy scales. This has
been interpreted as a signature of the so-called Luttinger liquid
behavior\cite{9,10}, characterized by the absence of electron
quasiparticles in the spectrum. Evidence of a power-law behavior
in the tunneling density of states at low energies has been
obtained from measurements in ropes\cite{11}, individual
single-walled nanotubes (SWNT)\cite{12} and multi-walled nanotubes
(MWNT)\cite{22}.

In the Luttinger liquid picture, critical exponents of observables
like the density of states are not universal and depend on the
interaction strength. In the case of carbon nanotubes, this refers
to the ubiquitous Coulomb interaction. However, a precise
determination of the interaction strength is precluded by the fact
that the electron-electron interaction is actually long-ranged,
and it can be assimilated to a coupling constant, only after
introducing a suitable infrared cutoff in the singular expression
of the one-dimensional (1D) Coulomb potential\cite{13,14}.

Furthermore, the shape of CN becomes relevant in setting the
strength of the electron correlations. In the case of MWNT, measurements
of the conductance refer usually to the outer layer, whose
electronic properties are influenced by the interaction with inner
metallic cylinders\cite{15}. Also, MWNT use to be significantly
doped, what leads to the presence of a large number of subbands at
the Fermi level\cite{25}. The contribution of a large number of modes
at low energies has then an appreciable impact in the enhancement of
observables like the tunneling density of states.

The purpose of the present paper is to study the combined effect
of the finite length of the electron system, on the one hand, and
of the number of subbands at low energies, on the other hand, in
the Luttinger liquid description of CN. With this aim, we adopt a
renormalization group (RG) approach, which is well-suited to
obtain the energy dependence of quantities like the quasiparticle
weight. In order to cure the infrared singularities arising from
the long-range interaction, we implement a dimensional
regularization of the theory, writing formally all bare quantities
slightly away from dimension $D = 1$\cite{24,ccm}. This formal
artifact has a real physical meaning, since the deviation $D - 1$
yields a measure of the finite size of the system, with the limit
$D \rightarrow 1$ corresponding to the case of increasingly long
nanotubes.

The Coulomb potential $1/|{\bf r}|$ can be represented in three
spatial dimensions as the Fourier transform of the propagator
$1/{\bf k}^2$. If the interaction is projected onto one spatial
dimension, by integrating for instance the modes in the two
transverse dimensions, then the Fourier transform has the usual
logarithmic dependence on the momentum, $V_C (k) = (1/2\pi ) \; \ln
(k_c /k) $. In this approach, $k_c$ is the memory that the
system keeps of the finite size of the transverse dimensions.
We choose instead to integrate formally a number $3 - D$ of
dimensions, so that the long-range potential gets the
representation
\begin{equation}
\frac{1}{|x|} =  \int \frac{d^D k}{(2\pi )^D} e^{ikx}
       \;   \frac{c(D)}{|{\bf k}|^{D-1}}
\end{equation}
where $c(D) = \Gamma ((D-1)/2)/(2\sqrt{\pi})^{3-D}$.
In the limit $D \rightarrow 1$, the expression of the potential
in momentum space becomes
\begin{equation}
\frac{c(D)}{|{\bf k}|^{D-1}} \approx \frac{1}{2\pi }
  \left( \frac{1}{D - 1} - \ln (|{\bf k}|) + \cdots  \right)
\label{corresp}
\end{equation}
This shows that $1/(D - 1)$ corresponds actually to the logarithm
of the high-energy cutoff $k_c$. In the RG framework, where all
quantities are scaled down to low momenta, the dimensionless
quantity $1/(D - 1)$ has to be traded for the logarithm of the
length of the system, measured in units of the nanotube finite
radius.

The great advantage of dealing with the above representation is
that the Coulomb interaction gives rise classically to a scale
invariant theory at any dimension above $D =1$, which allows
to proceed with the RG program. It is only at $D = 1$ that the
mentioned logarithmic dependence on the momentum leads to an
imperfect scaling behavior, making the use of the RG approach
not quite appropriate. Then, we write the
hamiltonian for the linear branches of CN in
the form
\begin{eqnarray}
H & = &  v_F \sum_{\alpha \sigma }
           \int_0^{\Lambda } d p |{\bf p}|^{D-1}
     \int  \frac{d\Omega }{(2\pi )^D} \;
   \Psi^{+}_{\alpha \sigma } ({\bf p}) \;
       \mbox{\boldmath $\sigma   \cdot $} {\bf p}
      \;     \Psi_{\alpha \sigma } ({\bf p})   \nonumber   \\
   \lefteqn{  +   e^2 \int_0^{\Lambda } d p |{\bf p}|^{D-1}
     \int  \frac{d\Omega }{(2\pi )^D}  \;
   \rho ({\bf p})   \;   \frac{c(D)}{|{\bf p}|^{D-1}}  \;
          \rho (-{\bf p})  \;\;\;\;\;\; }
\label{ham}
\end{eqnarray}
where the $\sigma_i $ matrices are defined formally by
$ \{ \sigma_i , \sigma_j \} = 2\delta_{ij}$ and $\rho ({\bf p})$
are density operators made of the electron fields
$\Psi_{\alpha \sigma } ({\bf p})$, with $\alpha $ labeling the
Fermi point and $\sigma $ the spin projection. The sum in Eq.
(\ref{ham}) runs then over the usual four modes present in
a single-walled nanotube, but it may include also the contribution
from a large number of subbands in the case of a doped multi-walled
nanotube.

We focus on the scaling properties of the model as the
cutoff $\Lambda $ is lowered, when a large number $N$ of subbands
contribute to the electronic properties down to the Fermi level.
Each subband is labeled with a different quantum number, which
corresponds to the momentum in the dimension around the nanotube.
For this reason, the dominant processes are those where
each scattered electron remains in the same linear branch.
The main effect of the interaction is to dress the bare electron
propagator with the polarization of the $N$ different subbands
given by
\begin{equation}
\Pi ({\bf k}, \omega_k) = 2 N b(D) \frac{v_F^{2-D} {\bf k}^2}
 { | v_F^2 {\bf k}^2 - \omega_k^2 |^{(3-D)/2} }
\label{pol}
\end{equation}
where $b(D) = \frac{2}{ \sqrt{\pi} } \frac{ \Gamma ( (D+1)/2 )^2
   \Gamma ( (3-D)/2 ) }{ (2\sqrt{\pi})^D \Gamma (D+1) }$ \cite{iz}.
The polarization (\ref{pol}) is the analytic continuation of the
known result for two linear branches with opposite chirality, which
we take away from $D = 1$ in order to carry out a consistent
regularization of the Coulomb interaction.
After dressing the interaction with the polarization (\ref{pol}),
the electron self-energy is given by the expression
\begin{eqnarray}
\Sigma ({\bf k}, \omega_k)  & = &  - e^2 \int_0^{\Lambda }
     d p |{\bf p}|^{D-1}  \int \frac{d\Omega }{(2\pi )^D}
    \int \frac{d \omega_p}{2\pi }     \nonumber   \\
 \lefteqn{   G ({\bf k} - {\bf p}, \omega_k - \omega_p)
 \frac{-i}{ \frac{|{\bf p}|^{D-1}}{c(D)} + e^2  \Pi ({\bf p},
    \omega_p) }     }
\label{selfe}
\end{eqnarray}

The low-energy properties of the theory are investigated by
taking the limit $\Lambda \rightarrow 0$, where the
self-energy $\Sigma $ turns out to have terms linear in $\omega_k$
and ${\bf k} $ that depend logarithmically on the cutoff.
This is the signal that the scale of the electron wavefunction
$Z^{1/2}$ and the Fermi velocity $v_F$ are renormalized at
low energies.
The divergent contributions to the electron propagator read
\begin{eqnarray}
\frac{1}{G}  & = &  \frac{1}{G_0} - \Sigma
     \approx  Z^{-1} ( \omega_k - v_F
  \mbox{\boldmath $\sigma \cdot$}{\bf k}) \;\;\;\;\;\;\;\;\;
        \;\;\;\;\;\;\;\;\;  \;\;\;\;\;\;\;\;\;    \nonumber  \\
 \lefteqn{   -  Z^{-1}   \frac{ f(D) }{2 N}
 \sum_{n=0}^{\infty} (-1)^n g^{n+1}    \left(
   \frac{n(3-D)}{n(3-D)+2}  \omega_k    \right.  }   \nonumber   \\
 \lefteqn{ +  \left.  \left(1 - \frac{2}{D} \frac{n(3-D)+1}{n(3-D)+2}
   \right)   v_F \mbox{\boldmath $\sigma \cdot$} {\bf k}
      \right)  h_n (D)   \log (\Lambda )   }
\label{prop}
\end{eqnarray}
where $g = 2 N b(D) c(D) e^2 / v_F $,
$h_n (D) = \frac{ \Gamma (n(3-D)/2 + 1/2) }
 { \Gamma (n(3-D)/2 + 1) }$, and
$f(D) = \frac{1}{ 2^D \pi^{(D+1)/2} \Gamma (D/2) b(D) }$ .

The usual RG argument is that the renormalized propagator $G$ must
be a finite quantity, so that the divergent dependences on the
cutoff $\Lambda $ have to be reabsorbed in the scale of the
wavefunction $Z^{1/2}$ and the Fermi velocity $v_F$ \cite{wen}. Under
a differential variation of $\Lambda $, $Z^{1/2}$ is renormalized
according to the equation
\begin{eqnarray}
\Lambda \frac{d}{d \Lambda} \log Z (\Lambda )  & = &
 -  \frac{ f(D) }{2 N} \sum_{n=0}^{\infty} (-1)^n g^{n+1}
     \;\;\;\;\;\;\;\;\;   \;\;\;\;\;\;\;\;\;    \nonumber  \\
  \lefteqn{  \frac{n(3-D)}{n(3-D)+2}   h_n (D)    }
\label{zflow}
\end{eqnarray}
The renormalization of $v_F$ can be translated into that of the
effective coupling $g = 2 N b(D) c(D) e^2 / v_F $, since the
electron charge $e$ is not renormalized in our model. The RG
equation for $g$ becomes
\begin{eqnarray}
\Lambda \frac{d}{d \Lambda} g (\Lambda )  & = &
  \frac{f(D)}{2 N} \frac{2(D-1)}{D}g^2 \sum_{n=0}^{\infty} (-g)^n
   \;\;\;\;\;\;\;\;\;   \;\;\;\;\;\;\;\;\;
             \;\;\;\;\;\;\;\;\;     \nonumber  \\
  \lefteqn{ \left(  \frac{(3-D)n+1}{(3-D)n+2}
 \right)  h_n (D)   }
\label{aflow}
\end{eqnarray}

We are now in a position to study the influence of the long-range
Coulomb interaction in the limit $D \rightarrow 1$. For this
purpose, we start by considering the RG equation (\ref{aflow}).
The series showing in the r.h.s. can be summed up
at $D = 1$, the flow equation taking then the following form in
the neighborhood of that point:
\begin{equation}
\Lambda \frac{d}{d \Lambda} g (\Lambda )
 \approx  \frac{1}{2 N} (D-1)
       g \left( 1 - \frac{1}{\sqrt{1+g}} \right)
\label{beta1}
\end{equation}
It follows from Eq. (\ref{beta1}) that, at $D = 1$, there is
formally a line of fixed-points covering all values of the
interaction strength. However, we still have to take into account
that the effective coupling $g(D)$ becomes singular in the
limit $D \rightarrow 1$. We can shuffle this divergence into
the initial value of the coupling, $g_0 (D)$, which turns out to
have near $D = 1$ the asymptotic behavior
\begin{equation}
  g_0(D) \approx  N \frac{ e^2} {  \pi^{2}v_F} \frac{1}{D-1}
\label{eqg0}
\end{equation}
Then, by matching the behavior of $g_0 (D)$ with that of Eq.
(\ref{beta1}), we observe that the $D - 1$ factor in the
r.h.s. of the RG equation is not completely canceled out
in the limit $D \rightarrow 1$ . This shows in a rigorous way that
the bare 1D long-range interaction is at a RG fixed-point for
arbitrary values of the interaction strength.

We consider next the RG equation (\ref{zflow}) for the
electron wavefunction scale in the limit $D \rightarrow 1$. The
series in the r.h.s. can be also summed up at $D = 1$,
with the result that the scaling equation in that limit reads
\begin{equation}
\Lambda \frac{d}{d \Lambda} \log Z (\Lambda )
\approx   \frac{1}{4 N}
 \left( \sqrt{1+g} + \frac{1}{\sqrt{1+g}} - 2 \right)
\label{anom}
\end{equation}
The function in the r.h.s. of Eq. (\ref{anom}) coincides with the
anomalous electron dimension found in the exact solution of the
Luttinger model\cite{emery,sol}.
This ensures that the RG approach is a sensible
way to obtain the low-energy properties of the model. We have to
bear in mind that our description introduces however two important
differences with respect to the usual treatment of 1D interacting
electrons. On the one hand, the slight deviation of the model from
$D = 1$ allows to control the approach to the bare long-range
interaction as the length of the system is increased. On the other
hand, we have also incorporated the effect of the number $N$ of
subbands that contribute at low energies, in order to account
for the influence of doping in MWNT.

With the RG approach we can face two different experimental
conditions, depending on the magnitude of the typical energy scale
involved in the measurement process, as compared to the spacing
between subbands in the nanotube sample. When the latter is larger
than the temperature or bias voltage applied to the sample, we are in
a situation where the number of subbands $N$ can be taken as constant
along the RG flow. Otherwise, for large enough temperature or bias
voltage, the number of subbands
that contribute at the scale of the high-energy cutoff is a
decreasing variable as $\Lambda \rightarrow 0$. We deal separately
with the two instances in what follows.

{\em RG approach with a constant number $N$ of subbands.---}
In transport experiments, the typical
scale of temperature or bias voltage lies usually below the
scale of the spacing between subbands. This has been so, even
in the measurements made in MWNT, where the spacing
corresponding to a typical diameter $d \approx 17 \; {\rm nm}$ is as
small as $\hbar v_F /d \approx 29 \; {\rm meV}$. In these conditions,
the only subbands that contribute to the properties measured
experimentally are those crossing the Fermi level.

We apply then Eqs. (\ref{eqg0}) and (\ref{anom}) to confront the
experimental results on the tunneling density of states gathered
from different nanotube samples. Starting with the measurements
made in SWNT, we take a number of subbands $N = 2$ in the
equations and adjust the deviation from $D = 1$ in accordance to
the length of the experimental sample. Following the argument
below Eq. (\ref{corresp}), we use the correspondence $1/(D - 1)
\approx \ln (L/d)$, $L$ being the nanotube length and $d$ the
nanotube diameter. A suitable choice corresponding to the
experiments reported in Ref.\cite{12} is $L/d \sim 10^3$, which
gives in turn $D  \approx 1.14$.

\begin{figure}

\par
\centering \epsfxsize= 7cm \epsfysize= 7cm
\epsfbox{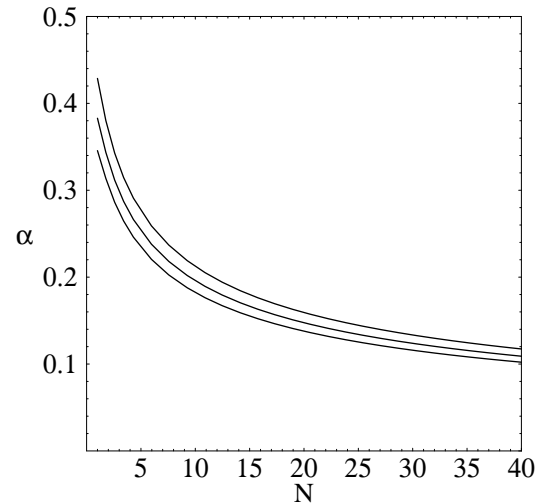}
\par

\caption{Estimates of the exponent $\alpha $ from the r.h.s. of
Eq. (\ref{anom}), for $e^2 /\pi^2 v_F  \approx 1.5 $ .
The different curves correspond, from top to
bottom, to $D = 1.14, 1.16$ and 1.18 .}
\label{one}
\end{figure}

The measurements of the tunneling density of states in SWNT have
shown a power-law dependence on energy, with values of the critical
exponent $\alpha $ accumulating around $\approx 0.35$ \cite{11,12}.
We have represented in Fig. \ref{one} the estimates obtained from the
r.h.s. of Eq. (\ref{anom}), for small values of $D - 1$. The
values of $\alpha $ have a smooth dependence on the length $L$ of
the system and fall around $\alpha \approx 0.35$ for $N = 2$ and
$D  \approx 1.14$, with a reasonable choice of the coupling $e^2
/\pi^2 v_F  \approx 1.5 $.

Our results show also an overall agreement with the exponents
measured in MWNT. It has been noticed that such systems are significantly
doped, so that a large number of subbands are found at the Fermi
level. The experimental conditions refer
to a situation where $N \approx 5 - 10$ (without taking into
account the spin degeneracy). It has been reported that the values
of the critical exponent $\alpha $ measured in 11 different samples
range from 0.24 to 0.37 \cite{22}.
This variation can be accounted for within
our RG approach by assuming that the number of subbands used in
the renormalization may shift from $N = 2$ to $N = 10$. Part of
the drift observed in the critical exponent may be also due to the
smaller aspect ratio of MWNT, although this
fact is difficult to assess given the lack of information about
the total length of the experimental samples.

We show then that the suppression of tunneling in
MWNT can be softened by increasing the doping
level. Our results may be also relevant in general for nanotubes of
large radius where there are a large number of subbands crossing
the Fermi level. These instances can be considered as midway in
the process of making contact with the physical properties of a
graphene sheet. This requires taking systems with larger
transverse size, what in turn may lead to a situation where the
spacing between subbands is smaller than the typical
energy scale in the experimental measurements. One has then to change
slightly the computational scheme, as discussed below.

{\em RG approach with a cutoff-dependent number of subbands
$N(\Lambda )$.---}
In samples of a very large diameter, one may envisage conditions
where the temperature or the bias voltage are much larger than
the subband spacing in the nanotube. The RG approach can be still
implemented, but taking into account that the number of subbands
contributing in the partial integration of modes at energy
$\Lambda $ depends on the value of the high-energy cutoff.
Let us suppose for simplicity that the system is at half-filling,
with the typical structure of two subbands crossing at the Fermi
level. For not too large energies, the number of subbands crossing
the energy level $\varepsilon $ has then
a linear dependence on $\varepsilon $, $N(\varepsilon ) = N_0 +
n_p \varepsilon $, where $N_0 = 2$ is the number of subbands at
the Fermi level.

\begin{figure}
\centering \epsfxsize= 7cm \epsfysize= 7.45cm
\epsfbox{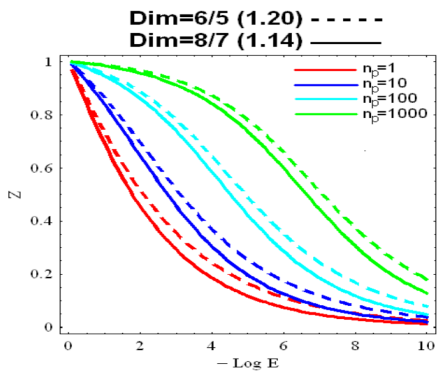}
\caption{Energy dependence of the quasiparticle
weight $Z$ at dimensions 6/5 and 8/7, for different values of
$n_p$.}
\label{two}
\end{figure}

We obtain the scaling of the quasiparticle weight $Z$ and
the different observables by imposing a dependence of $N$ on the
cutoff $\Lambda $ according to the above formula. This description
yields a sensible prediction for
experiments where the average effect of a large
number of subbands is measured. The dependence of the quasiparticle
weight $Z$ on energy as the cutoff is sent towards the Fermi level
is represented in Fig. \ref{two}, for different values of $n_p$.
In the present instance, the dominant contribution to the power-law
behavior of the tunneling density of states comes from
the number of subbands varying with the energy scale.
This sets the value of $\alpha $ close to 1 from the start, what is
the natural way of recovering the characteristic linear density of
states of a graphite layer from our 1D point of view.

We observe from the results in Fig. \ref{two} that the
quasiparticle weight $Z$ tends to have a flat behavior at high
energies for large values of the density of subbands $n_p$. This is
in contrast to the rapid decrease signaling the typical power-law
behavior for small values of $n_p$. In the curves for $n_p = 100$
and $n_p = 1000$, we see the existence of a crossover between a
regime with persistent quasiparticles and another characteristic
of the Luttinger liquid behavior. The physical interpretation is
that, for high energies above the crossover scale, the system has
similar properties to the 2D graphene, while one has to look at
sufficiently small energy scales (or large length scales) to measure
the properties of the 1D wire.

We have obtained several results that may be checked against future
measurements carried out in MWNT and nanotubes of very large
radius. We predict that the exponent giving the power-law behavior
of the tunneling density of states may suffer a significant
reduction upon doping those systems, with the possibility of
reaching values as small as 0.1 for $N \approx 40$. We have also shown
that, when dealing with nanotubes of very large radius, there is a
high-energy regime with persistent electron quasiparticles which
has properties closer to 2D graphene than to the Luttinger liquid.
We believe such features may be of interest when developing
carbon-based devices made of graphene and nanotube structures with
different shapes.\vspace{.25cm}

 This work is partly supported by
the Italian Research Ministry, National Interest Program.

\end{document}